# Surface Pourbaix plots of M@N₄-graphene single-atom electrocatalysts from Density Functional Theory thermodynamic modelling

Ana S. Dobrota[1], Natalia V. Skorodumova[2], Slavko V. Mentus[1,3], Igor A. Pašti[1,2]*

[1]*University of Belgrade – Faculty of Physical Chemistry, Belgrade, Serbia*

[2]*Department of Materials Science and Engineering, School of Industrial Engineering and Management, KTH – Royal Institute of Technology, Stockholm, Sweden*

[3]*Serbian Academy of Sciences and Arts, Belgrade, Serbia*

*Correspondence: igor@ffh.bg.ac.rs;

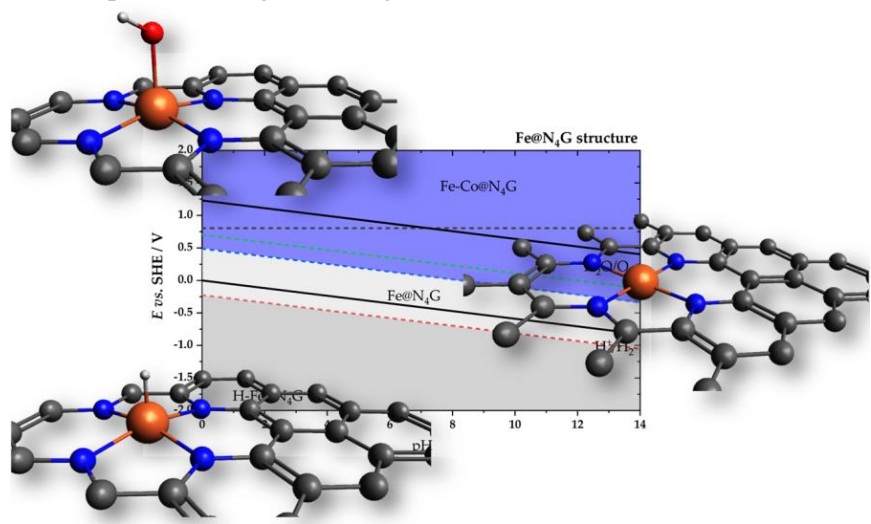

**Abstract**

Single-atom catalysts (SACs) are rapidly developing in various application areas, including electrocatalysis of different reactions, usually taking place under harsh pH/electrode potential conditions. Thus, a full atomic-level understanding of the nature of the active sites under realistic electrochemical conditions is needed, having in mind that the state of SACs active centers could be altered by the adsorption of spectating species. In this contribution, Density Functional Theory is employed to conduct thermodynamic analysis of SACs with metal atoms (Mn, Fe, Co, Ni, Cu, Ru, Rh, Pd, Ag, Ir, Pt, or Au) embedded into $N_4$ moiety in graphene. Various surface electrochemical processes on such SACs are considered, their Pourbaix plots are constructed, and their activity, selectivity, and stability under operating conditions are discussed. It is demonstrated how adsorption of H, O and OH can cause blockage and restructuring of the active sites and alter the electronic structure. Furthermore, when one deals with metals with lower *d*-band filling, it is shown that metal center oxidation is preferred over the oxidation of carbon lattice. The effect of the state of the metal center on the reactivity of the carbon lattice is discussed in the case of Fe@$N_4$-graphene. Finally, a possible strategy for confirming the changes in the architecture of the SACs' active site by analyzing their vibration spectra is suggested.

# 1. Introduction

The battle for a sustainable energy society is closely related to the effectiveness of the energy conversion process and technologies, as well as the rationalization and economization of many industrial processes crucially connected to energy conversion. The core of most of these processes is associated with different kinds of catalytic reactions, where a suitable catalyst plays a key role. The race for active and stable catalysts is also connected to developing low-cost catalysts with high utilization of active materials. Probably the youngest addition to the catalyst world is Single-Atom Catalysts (SACs). In this class of catalysts, the active centers are constructed of single atoms embedded into a properly designed matrix whose contribution to the catalyst performance is crucially connected to the active atom centers [1–3], while typical activity trends established for bulk or nanosized catalysts are easily broken [4].

The rapid development of the SACs field started one decade ago [5], and in the last two years, the annual number of related publications on SACs exceeded a couple of thousands (according to the Scopus search). Besides being attractive due to an ultimate level of catalyst utilization, SACs also provide a possibility for fundamental analysis of the mechanisms of catalytic processes if the identical catalytic sites are dispersed in the sample [1]. However, the characterization of SACs is not routine as single-atom sites have to be confirmed. This requires high-resolution (i.e., atomic resolution) electron microscopy and synchrotron-based techniques, like X-ray absorption near edge structure (XANES), which is employed to resolve the coordination of active sites [6,7]. However, from the theoretical point of view, SACs are absolutely ideal systems to investigate. While in the case of bulk or nanosized catalysts, the models have to be set with a certain level of simplification, in the case of SACs, theoretical models could easily capture the realistic architecture of catalytically active sites even with modest computational costs. Thus, a combination of atomic-level characterization and electronic structure-level calculations (mostly based on Density Functional Theory, DFT) does seem like a perfect way to a fundamental understanding of the operation and underlying principles governing SACs activity, selectivity and stability.

Many SACs have found their applications in electrocatalysis and are being investigated for different reactions, like hydrogen evolution reaction (HER) [8–10], oxygen reduction reaction (ORR) [11–14], and others discussed further in this work [15–19]. However, a general characteristic for all electrocatalytic reactions is that they occur under rather harsh conditions – very low or very high pH and extreme electrode potentials, at least in some parts of the potential window where a given reaction takes place. Interestingly, in the case of extended electrocatalyst surfaces and nanoparticles, surface electrochemical processes have been unambiguously linked to the catalytic activity [20–22], but this is not the case with SACs. In fact, a literature survey shows that SACs are always considered perfect, without a possibility of active centers getting covered by any of spectating species. Such adsorption processes could arise due to the exposure of SACs to potentials and pH values where such spectating species adsorbed on active sites could be thermodynamically stable. However, there are examples of *in operando* studies where it was explicitly confirmed that the nature of active sites in SACs changes with the operating potential [23]. Moreover, our recent theoretical work showed that single metal atoms (Ru, Rh, Ir, Ni, Pd, Pt, Cu, Ag, and Au) embedded into graphene monovacancy should always be covered either by $H_{ads}$ or oxygen species ($OH_{ads}$ or $O_{ads}$) depending on the pH and electrode potential [24].

In this work, we continue previous DFT-based thermodynamic analysis of SACs focused on constructing Pourbaix plots for SACs, which connect the electrode potential and pH domains of thermodynamical stability of different phases under electrochemical conditions [25,26]. Examples of Pourbaix plots construction for SACs can be found in the literature [27,28]. However, although this concept provides valuable insights into the stability of different phases under electrochemical conditions, it is not widely used elsewhere. In this work, we focus on the metal atoms (Mn, Fe, Co, Ni, Cu, Ru, Rh, Pd, Ag, Ir, Pt, and Au) embedded into $N_4$ moiety in graphene lattice, which is one of the most intensely investigated types of SACs for different electrochemical reactions [29,30], although other nitrogen coordination environments of metal centers are also frequently reported [31–33]. In particular, we consider the following surface electrochemical process on model M@$N_4$-graphene SACs:

$$M^{z+} + ze^- + vG \rightarrow M@N_4\text{-graphene}, \qquad (1)$$

$$M@N_4\text{-graphene} + H^+ + e^- \rightarrow H-M@N_4\text{-graphene}, \qquad (2)$$

$$OH-M@N_4\text{-graphene} + H^+ + e^- \rightarrow M@N_4\text{-graphene} + H_2O, \qquad (3)$$

$$O-M@N_4\text{-graphene} + 2H^+ + 2e^- \rightarrow M@N_4\text{-graphene} + H_2O. \qquad (4)$$

Additionally, we identify the pH *vs*. potential regions of different surface phases. Based on the obtained results, we discuss the activity, selectivity, and stability of studied SACs under operating electrochemical conditions and suggest possible strategies for identifying active sites restructuring. Our results suggest that special care is needed when interpreting the results of *ex-situ* characterization and modelling of SACs when linking the results of such analyses to the experimental data related to electrochemical performance.

## 2. Methods

Pristine graphene was modelled using a 4×4 cell (32 C atoms, $C_{32}$), which was confirmed to be large enough for the purposes of this study [34,35]. Graphene with an $N_4$ defect ($N_4G$) was obtained by removing two adjacent C atoms from $C_{32}$ and substituting their first C neighbors with N, resulting in the $C_{30}N_4$ model. Metal-doped $N_4G$ systems ($MN_4G$) were constructed by embedding metal atoms into the $N_4$-vacancy site of $N_4G$, at the center of 4 N atoms, and relaxing the structures.

The first-principle DFT calculations were performed using the Vienna *ab initio* simulation code (VASP) [36–38]. The Generalized Gradient Approximation (GGA) in the parametrization by Perdew, Burk and Ernzerhof [39] combined with the projector augmented wave (PAW) method was used [40]. The cut-off energy was 600 eV, while Gaussian smearing with a width of $\sigma = 0.025$ eV for the occupation of the electronic levels was used. A Monkhorst–Pack $\Gamma$-centered 10×10×1 $k$-point mesh was used. During the optimization, all atoms in the simulation cell were allowed to relax until the Hellmann–Feynman forces acting on them became smaller than $10^{-2}$ eV Å$^{-1}$. Spin polarization was included in all calculations. DFT+D3 approach of Grimme[41], which corrected the total energy by a pairwise term, was used to include the dispersion interactions.

The energy of the metal atom embedding into the $N_4$-vacancy site of $N_4G$ is quantified by its embedding energy:

$$E_{emb}(M) = E_0[MN_4G] − E_0[N_4G] − E_0[M], \qquad (5)$$

where $E_0$ is the ground state energy of the metal adsorbed onto $N_4$-vacancy site of $N_4$-graphene [$MN_4G$], $N_4$-graphene with a vacancy [$N_4G$] and isolated metal atom [M]. After obtaining $MN_4G$ systems by M embedding into $N_4G$ their reactivity is probed using H, O and OH as adsorbates. We have investigated all symmetrically nonequivalent sites of studied SACs, as shown in **Figure 1**. The reactivity of $MN_4G$ systems is described by the magnitude of the adsorption energies of these species, calculated analogously to $E_{emb}(M)$:

$$E_{ads}(A) = E_0[A−MN_4G] − E_0[MN_4G] − E_0[A], \qquad (6)$$

where $E_0$ is the ground state energy of the adsorbate A (A = H, O or OH) adsorbed on $MN_4G$ [$A−MN_4G$], $MN_4G$ system [$MN_4G$], and isolated adsorbate [A]. Negative $E_{ads}$ value indicates exothermic adsorption.

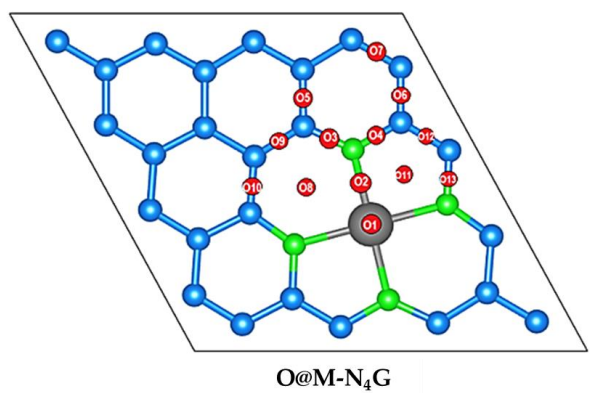

**O@M-N$_4$G**

**Figure 1.** Investigated nonequivalent adsorption sites (O1-O13) for H, OH, and O on studied SACs.

Vibrational analysis was done using a second-order finite differences approach with a displacement of 0.015 Å in all three directions. Constrained dynamics was employed so that vibrational frequencies were calculated taking the displacements of adsorbates and active SAC site up to the second nearest neighbors of metal sites. This significantly reduces the computational costs while providing a satisfactory accuracy. Namely, using the constrained dynamics where only adsorbates vibrations are considered, the Gibbs energies are underestimated by 0.03 to 0.06 eV [42]. Using thus obtained vibrational frequencies, the zero-point energies (ZPE) and vibrational contributions to the entropy ($TS_{vib}$) were calculated. Standard potentials (*vs.* Standard Hydrogen Electrode, SHE) for reactions (2)–(4) were calculated using the Computational Hydrogen electrode approach [43]. Assuming the equilibrium of hydrogen electrode:

$$H^+ + e^- \leftrightarrow H_{2(g)}, \tag{5}$$

the electrochemical potential of $(H^+ + e^-)$ was matched to that of $H_2$ at pH = 0. Then, for each of the competing phase $i$ in reactions (2)–(4), the chemical potential $(\mu_i)$ was calculated as

$$\mu_i = E_{tot} + ZPE - TS_{vib}, \tag{6}$$

The electric field effects, acting through the dipole moments, were disregarded, as justified in ref. [44]. To obtain the chemical potential of liquid water, we calculated its chemical potential at the gas phase at 298 K and 1 atm and then corrected it by Gibbs free energy change ($\Delta G$) for evaporation under the same conditions. The equilibrium potentials for reactions (2)-(4) were calculated by taking the given reaction (2)-(4) as a cathode in a hypothetical galvanic cell with a hydrogen electrode as the anode, similarly to the approach used in ref. [45]. Gibbs free energy changes ($\Delta G$) were calculated for each reaction as:

$$\Delta G = \sum_{i, products} \mu_i - \sum_{i, reactants} \mu_i, \tag{7}$$

The electromotive force of a hypothetical galvanic cell is obtained by dividing calculated $\Delta G$ (in eV) by the number of electrons exchanged in the reaction. Taking that the anode is SHE ($E° = 0$ V), the obtained values are numerically equal to the standard electrode potentials for reactions (2)-(4). For the reaction given by Eq. (1) metal dissolution potential from the $N_4$-graphene site was calculated as in ref. [46], considering a hypothetical galvanic cell where one electrode is a massive piece of metal M, while the other is M@vG electrode. For the full description, the reader is referred to ref. [46]. For the construction of the surface Pourbaix plots, the activity ($a$) of $M^{z+}$ ions was taken to be $1 \times 10^{-8}$ mol dm$^{-3}$.

For selected systems we addressed the reactivity of $N_4$-graphene basal plane using Fukui indices [47,48]. These indices can be used to assess the affinity of particular atom A in a molecular system with $N$ electrons in its ground state towards nucleophilic ($f^+$), electrophilic ($f^-$), or radical attack ($f^0$), and are evaluated as:

$$f^+_A = P_A(N+1) - P_A(N) \tag{8}$$
$$f^-_A = P_A(N) - P_A(N-1) \tag{9}$$
$$f^0_A = (P_A(N+1) - P_A(N-1))/2 \tag{10}$$

where $P_A(N+1)$, $P_A(N)$, and $P_A(N-1)$ stand for the electron population of atom A in a molecular system with $N+1$, $N$, and $N-1$ electrons, respectively. It is clear that Fukui indices are dependent on the computational scheme used to evaluate population, and here we used the Bader approach [49].

## 3. Results and Discussion

First, we address metal embedding into the $N_4$-graphene site and adsorption of studied adsorbates (H, OH, and O). As shown in **Figure 2** (left), within the considered SACs there are strong interactions of metal SACs with $N_4$-graphene moieties, surpassing cohesive energies of corresponding bulk metal phases, except for Ag and Au. This observation is the first indication of enhanced thermodynamic stability of metal centers towards dissolution.

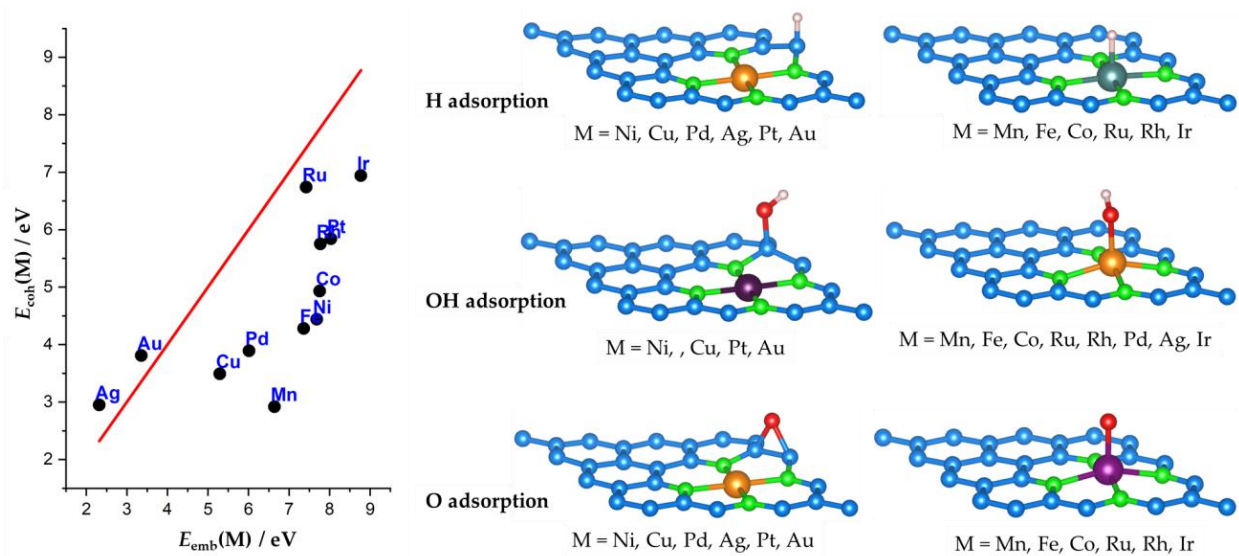

**Figure 2.** The parity plot of cohesive *vs*. embedding energies of studied metal single atoms (left) and classification of studied SACs according to preferential adsorption sites towards $H_{ads}$, $OH_{ads}$ and $O_{ads}$ (right). The red line in the left diagram is 1-1 line, and it separates the positions of atoms with low and high energies of interaction with $N_4$-graphene moieties, when compared to cohesive energies of bulk metals.

Studied models SACs differ greatly in terms of their reactivity of metal center and surrounding $N_4$-graphene matrix (**Figure 2**, **Table 1**). Upon the systematic analysis of H, OH, and O adsorption (see sketchs in right part of Fig. 2), we find that only 6 model SACs preferentially bind all these three adsorbates at the metal sites. These are the ones with Mn, Fe, Co, Ru, Rh, and Ir metal centers. This result draws a very clean line separating $d^n$ elements into two groups: to those with the lower occupation of d orbitals (n ≤ 8) for which the reactivity (measured as adsorption energy) is concentrated at the metal center, and to those with the higher occupation of *d* orbitals (n > 8). For the second group, metal centers are either never preferential for studied adsorbates (Pt and Au), or the situation depends on the given adsorbate (Ni, Cu, Pd, Ag, **Figure 2**). Looking at the overall trends along the periods of the PTE, the lowest reactivity for all the studied model SACs is seen for those with Group 10 metals (Ni, Pd, Pt, **Table 1**), where C atoms bonded to N are found to be most reactive. For these SACs, we believe that metal centers are not (re)active ones but actually graphene lattice surrounding $MN_4$ moiety. Among studied systems, we also observed some peculiarities. First, Ru is very oxophilic, and OH adsorbed at the $RuN_4$ site is found to be dissociated. Second, when OH binds to $AgN_4$, spontaneous detachment of AgOH molecule is seen. This is in line with very weak Ag bonding to the $N_4$ site (**Figure 2**). However, stable OH bonding to the carbon matrix is seen with $E_{ads}$ = −1.68 eV.

**Table 1**. Adsorption energies of H, OH, and O on studied model SACs

| Periode | Metal | $E_{ads}(H)$ / eV | $E_{ads}(OH)$ / eV | $E_{ads}(O)$ / eV |
|---|---|---|---|---|
| **3d** | Mn | −2.06 | −2.96 | −4.54 |
| | Fe | −2.26 | −2.88 | −4.14 |
| | Co | −2.30 | −2.34 | −2.83 |
| | Ni | −1.35 | −1.43 | −2.57 |
| | Cu | −1.37 | −1.57 | −2.63 |
| **4d** | Ru | −2.70 | −3.72* | −4.84 |
| | Rh | −2.68 | −2.35 | −2.88 |
| | Pd | −1.37 | −1.04 | −2.62 |
| | Ag | −1.55 | −1.95** | −2.92 |
| **5d** | Ir | −2.82 | −2.32 | −3.17 |
| | Pt | −1.38 | −1.43 | −2.61 |
| | Au | −1.51 | −1.76 | −3.05 |

* dissociated OH group, **AgOH spontaneous detachment.

Using the ground state energies, ZPE, and entropic contributions for the M@N$_4$-graphene, H−M@M@N$_4$-graphene, HO−M@M@N$_4$-graphene, and O−M@M@N$_4$-graphene systems, we constructed surface Pourbaix plots for the studied model SACs at standard conditions and at 298 K. Considered surface processes include:

(i) metal dissolution (Eq. (1)) with Nernst equation:
   $E(M^{z+}/M@N_4G) = E°(M^{z+}/M@N_4G) − (0.059/z)×\log a(M^{z+})$,   (11)

(ii) hydrogen deposition (analogous to hydrogen underpotential deposition (UPD) on extended metal surfaces [50–52], Eq. (2)), with Nernst equation:
   $E(M@N_4G/H−M@N_4G) = E°(M@N_4G/H−M@N_4G) − 0.059×pH$,   (12)

(iii) SACs oxidation *via* deposition of OH$_{ads}$, Eq. (3), with Nernst equation:
   $E(OH−M@N_4G/M@N_4G) = E°(OH−M@N_4G/M@N_4G) − 0.059×pH$,   (13)

(iv) SACs oxidation *via* deposition of O$_{ads}$, Eq. (4), with Nernst equation:
   $E(O−M@N_4G/M@N_4G) = E°(O−M@N_4G/M@N_4G) − 0.059×pH$.   (14)

The calculated standard potentials ($E°(O/R)$) are summarized in **Table 2**. Metal dissolution, Eq. (1) is not pH-dependent, but H$_{ads}$, OH$_{ads}$, and O$_{ads}$ formation are, with the slope of the equilibrium potential versus pH line of 0.059 mV *per* pH unit in all the cases. When the equilibrium potentials were calculated using Eqs. (11)-(14) for pH ranging from 0 to 14, the stable phases were identified following the rule that the most stable oxidized phase had the lowest equilibrium potential, while the most stable reduced phase was the one with the highest equilibrium potential.

Some general rules can be outlined by analyzing the overall trends in the calculated standard potentials for considered reactions. First, the stability of metal sites towards dissolution increases along the periods, and it drops for coinage metals (Cu, Ag, Au). This trend generally follows the trend in dissolution potentials of bulk metal phases, and the drop observed for coinage metals is due to very low embedding energies into the N$_4$-graphene moiety (**Figure 2**, left). Second, the affinity of M@N$_4$-graphene SACs towards H$_{ads}$ decreases along the period of the PTE and H$_{ads}$ formation starts at progressively more negative potentials. Along with the reduced affinity towards H$_{ads}$, the oxophilicity of M@N$_4$-graphene SACs also decreases. In other words, the potentials for OH$_{ads}$ and O$_{ads}$ deposition shift to higher potentials, and these two potentials scale rather well. Such scaling is the consequence of the scaling between OH and O adsorption energies (**Table 1**) and the relatively constant contribution of ZPE and $TS_{vib}$ terms which weakly depend on the adsorption site (in the overview in **Table 2**, we did not differentiate between adsorption at metal sites and carbon lattice).

**Table 2**. Calculated standard potentials for reactions given by Eq. (1) to (4), standard conditions, 298 K, pH = 0.

|    | z | $E^0$(M$^{z+}$/M@N$_4$G) / V | $E^0$(H-M@N$_4$G/M@N$_4$G) / V | $E^0$(OH-M@N4G/M@ N$_4$G) / V | $E^0$(O-M@ N$_4$G /M@ N$_4$G) / V |
|----|---|------|--------|--------|--------|
| Mn | 2 | 0.627 | −0.401 | 0.484 | 0.520 |
| Fe | 2 | 1.040 | −0.224 | 0.486 | 0.704 |
| Co | 2 | 1.056 | −0.608 | 1.045 | 1.333 |
| Ni | 2 | 1.322 | −1.432 | 2.194 | 1.692 |
| Cu | 2 | 1.192 | −1.391 | 1.779 | 1.642 |
| Ru | 3 | 0.798 | 0.392 | −0.508 | 0.357 |
| Rh | 3 | 1.401 | 0.163 | 1.045 | 1.358 |
| Pd | 2 | 1.919 | −1.386 | 2.303 | 1.657 |
| Ag | 1 | 0.160 | −1.284 | 1.929 | 1.524 |
| Ir | 3 | 1.727 | 0.316 | 1.075 | 1.190 |
| Pt | 2 | 2.241 | −1.396 | 2.230 | 1.674 |
| Au | 3 | 1.345 | −1.999 | 1.815 | 1.438 |

First, we consider M@N$_4$-graphene catalyst where M is Mn, Fe, or Co (**Figure 3**). These metals are located in the middle of the 3d series, and the corresponding catalysts are commonly reported as highly active ones for ORR, in particular, Fe-based ones, for which experimental data in both acidic and alkaline media have been reported [23,53]. Our results suggest that Mn and Co SACs are expected to dissolve in acidic media (just below pH = 2 for Mn, and pH = 4 for Co), while the Fe center is stable at any pH. The metal sites are bare until 0.484 V and 0.486 V (**Table 2**) for Mn and Fe, while for Co, the formation of OH$_{ads}$ at the metal site is expected at 1.045 V *vs*. RHE. From the low potential side, H$_{ads}$ formation is not favored at least until −0.224 V *vs*. RHE (in the case of Fe center), and in current computational models for HER, this potential is essentially equalized to the HER overpotential [54]. However, when analyzing current theoretical models for ORR on these materials, it is generally observed that the possibility of O$_2$ competing with other adsorbates at metal centers is disregarded. We note that we do not consider additional anion adsorption here, but the discussion of the effects of anions can be found in ref. [55]. The authors showed that anion adsorption could actually render higher ORR activity of metal centers by adjusting Gibbs energies of formation of different ORR intermediates.

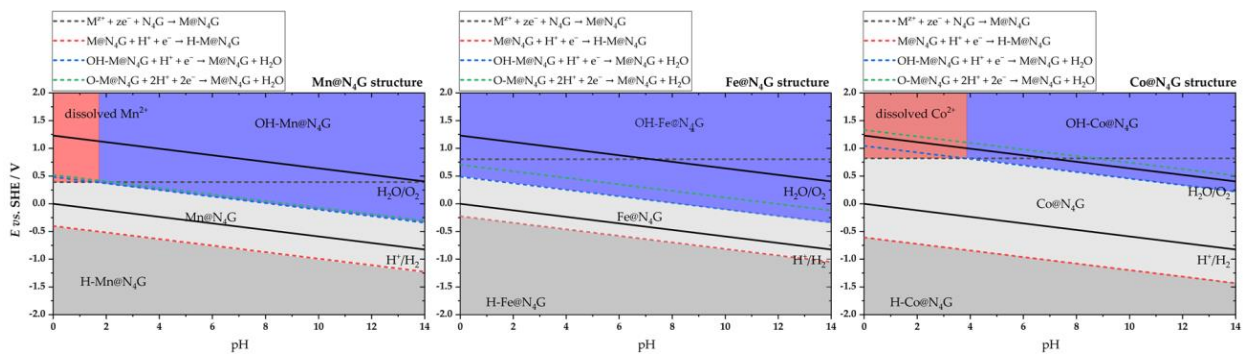

**Figure 3**. Surface Pourbaix plots for Mn@N$_4$-graphene, Fe@N$_4$-graphene, and Co@N$_4$-graphene model SACs. Thick black lines indicate the theoretical water stability region.

While the redox transitions of M-N$_4$ centers in the ORR operating potential region are well known and even used as activity descriptors [28], this is typically not considered when modelling ORR on this class of materials. However, the existence of OH$_{ads}$ phase on Fe-N$_x$ centers has been confirmed experimentally as well [23]. In fact, *in operando* study of Jia *et al*. [23] showed that the redox switching of Fe-N centers, corresponding to Fe$^{2+}$/Fe$^{3+}$ transition and its dynamic nature, is the key for high ORR activity. Specifically, using XANES and FT-EXAFS, the mentioned work showed OH$_{ads}$ formation on Fe-N centers at potentials corresponding to ORR, particularly those close to the ORR onset (around 0.9 V *vs*. RHE). Moreover, for Fe-N$_4$ catalysts, it was recently shown that different active sites are operative for ORR in acidic and alkaline media [56]. The authors demonstrated that in alkaline media, Fe-N$_4$ sites do not contribute ORR as being blocked with two OH$^-$, and that ORR activity is due to graphitic N sites.

Our findings of the existence of OH$_{ads}$ at the M centers in the ORR operating region could also be correlated with the higher H$_2$O$_2$ yield in the potential region 0.4 to 0.9 V *vs*. RHE on Mn-N and Fe-N catalysts [57], which is behavior opposite to platinum, where H$_2$O$_2$ yield increase in the low potential region where hydrogen underpotential deposition starts. Finally, speaking about the ORR selectivity of these catalysts, Fe-based catalysts are typically reported to have high selectivity towards 4e reduction, while Co-based ones reduce O$_2$ to peroxide [58,59]. However, the H$_2$O$_2$ yield is also affected by catalyst loading [60] and probably determined not only by the intrinsic activity and selectivity of active sites.

Next, we consider Ru, Rh, and Ir-based M@N$_4$-graphene catalysts (**Figure 4**). Ru@N$_4$-graphene SAC exhibits specific behavior as compared to other systems considered here. H$_{ads}$ is strongly bound to the Ru site, while this site is also very oxophilic, so the Ru site is either covered with hydrogen or oxygen (**Figure 4**). High oxophilicity of the Ru site agrees with the results of the computation screening of SACs for ORR [61], where also Rh@N$_4$-graphene and Ir@N$_4$-graphene are outlined as the most active for ORR, along with its Co-based analog. Separate theoretical work has also confirmed an expectedly high ORR activity of Ir@N$_4$-graphene SAC [62]. Here we show that Rh@N$_4$-graphene and Ir@N$_4$-graphene catalysts undergo OH$_{ads}$ formation at potentials above 1 V *vs*. RHE, suggesting these metal sites could indeed be free of competing phases under ORR conditions. In this sense, Rh@N$_4$-graphene and Ir@N$_4$-graphene

behave similarly to Co-based SAC, but the differences are observed at potentials close to the hydrogen evolution region. While for Co-based SAC hydrogen deposition starts at potentials below −0.6 V *vs*. RHE, for Rh and Ir-based SACs this process takes place at potentials positive to that of 0 V *vs*. RHE. In electrochemical words, hydrogen underpotential deposition is expected in these cases, just like in the cases of extended metallic surfaces [50,52]. As discussed previously, the potential of $H_{ads}$ formation could be equalized to the overpotential for HER, when using the approach of Nørskov *et al.* [54] to estimate HER activity. This also puts Rh@$N_4$-graphene and Ir@$N_4$-graphene at the top of the HER activity list among SACs studied here.

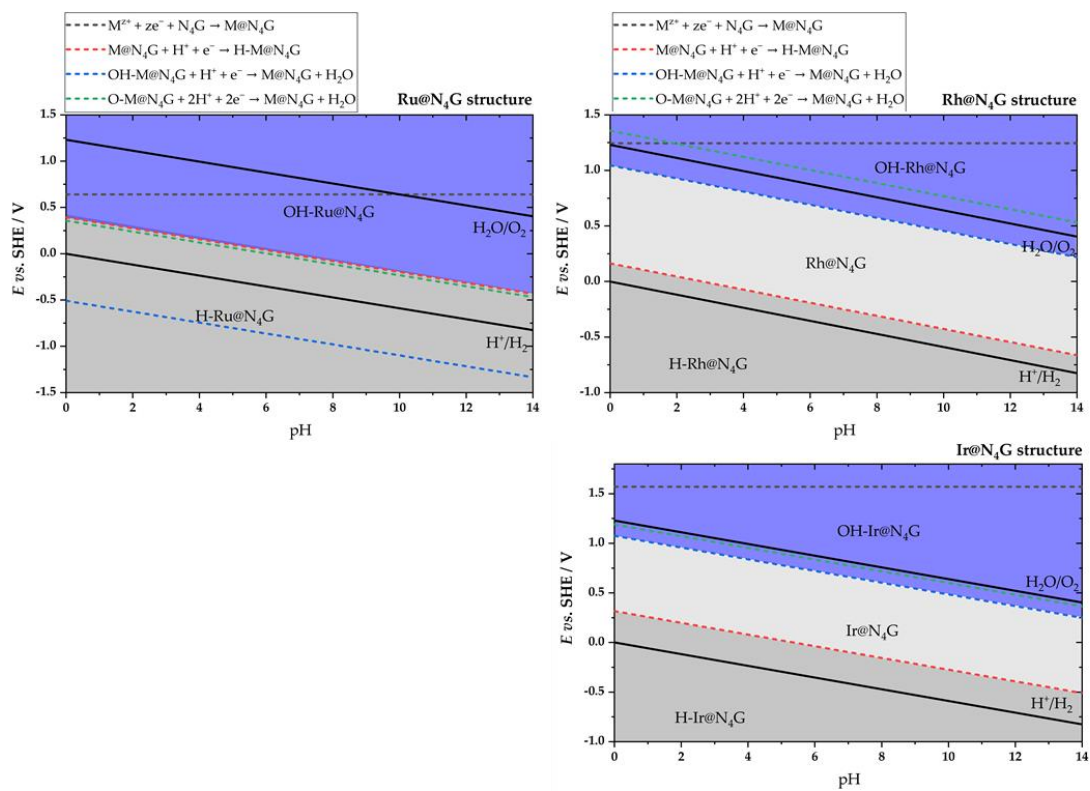

**Figure 4**. Surface Pourbaix plots for Ru@$N_4$-graphene, Rh@$N_4$-graphene, and Ir@$N_4$-graphene model SACs. Thick black lines indicate the theoretical water stability region.

However, if one considers strong $H_{ads}$ bonding to Rh- and Ir-centers, it is necessary to consider the adsorption of the second H at the metal site. In this model, first $H_{ads}$ would strongly bind hydrogen, and HER would proceed *via* second adsorbed hydrogen. However, we note that this could only be possible if M-$N_4$ moiety is embedded in graphene but not for grafted M-$N_4$ moieties (in the latter case, only one site of M-$N_4$ is available for interaction). When the adsorption of the second H is considered, an average H adsorption energy (calculated *vs*. $H_2$ molecule) is 0.23 eV for Rh-, and 0.00 eV for Ir-based SAC. Using the model from ref. [54] to estimate HER activity, this actually means that for Rh, the activity would be the same as estimated from the adsorption of one H, but the position of this SAC on HER volcano would be different (shift from strong adsorbing branch to weak adsorbing branch). However, the activity of Ir-based SAC would be higher, as the first H$_{ads}$ rendered the adsorption energy of the second H$_{ads}$ to a more favorable value, causing the shift of this SAC closer to the apex of HER volcano.

Finally, the model SACs with metal centers from group 10 (Ni, Pd, and Pt) and group 11 (Cu, Ag, and Au) generally show poor affinity towards H$_{ads}$ and the considered oxygen species (**Figure 5**). Moreover, the Ni-, Cu-, Ag-, and Au-based SACs are prone to dissolution at low pH, while all these metal centers tend to be covered with O$_{ads}$ rather than OH$_{ads}$. As a result of low reactivity, these metal centers are indeed bare in the water stability region (if not dissolved) and possess poor catalytic activity, as previously found for ORR and OER on Ni-, Pd-, and Pt-based M-N$_4$ SACs [61].

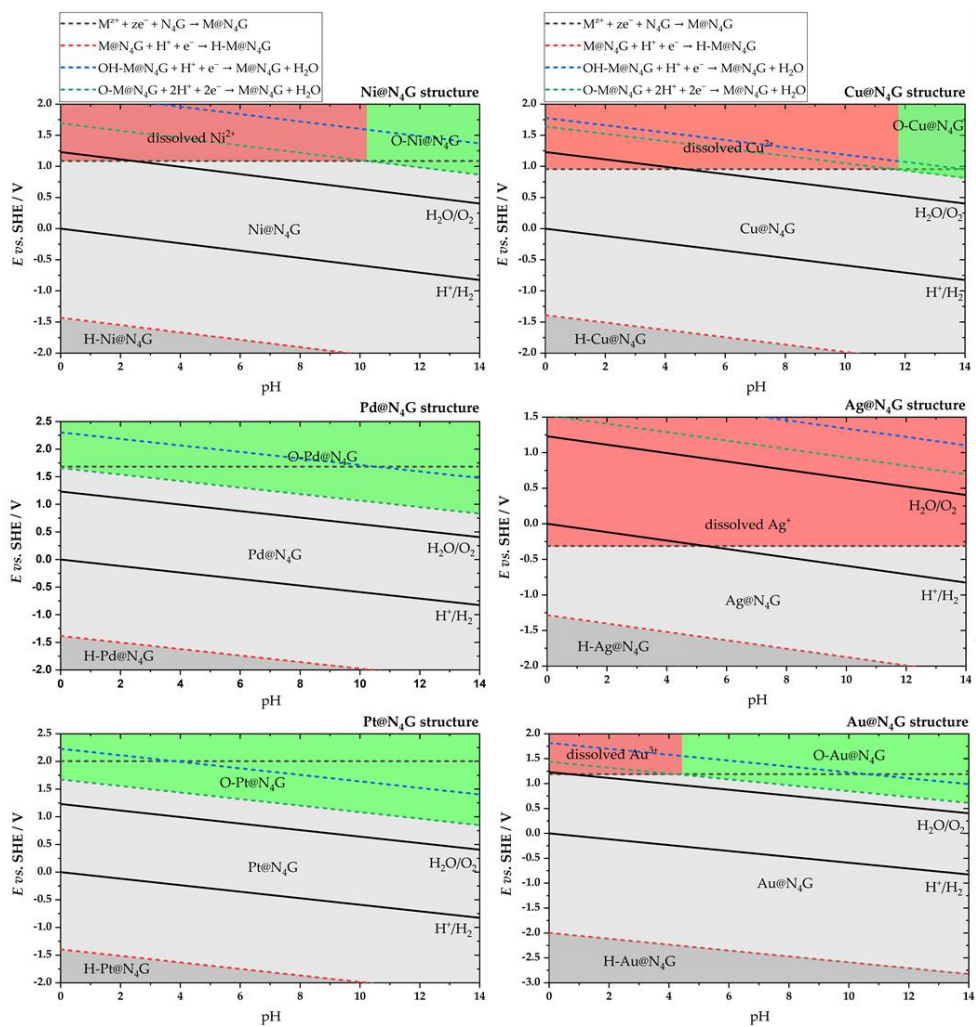

**Figure 5**. Surface Pourbaix plots for Group 10 model SACs (Ni@N$_4$-graphene, Pd@N$_4$-graphene, and Pt@N$_4$-graphene) and Group 11 model SACs (Cu@N$_4$-graphene, Ag@N$_4$-graphene, and Au@N$_4$-graphene). Thick black lines indicate the theoretical water stability region.

While we mainly discussed the impact of different surface phases of SACs on ORR and HER, the situation is more complex as SACs are nowadays applied for a range of electrocatalytic reactions (**Figure 6**) [1,15,63]. For bulk metallic surfaces and nanoparticles, it is well established that the spectating species (ions, adsorbed spectators) mainly render the catalytic activity *via* surface coverage. However, for SACs, the situation is different as the adsorption of spectating species and electrocatalytic processes occur at the same site. Hence, the effect is dual, as the site blockage goes along with the alteration of the electronic structure (**Figure 6**). Particularly in the case of $M@N_4$ catalysts, it is important to emphasize that spectator adsorption can have a detrimental effect on the catalytic activity if the moiety is grafted on the carbon surface rather than embedded in graphene leaving only one side of the active center to the reacting phase. For the case of $Fe@N_4$-graphene catalyst, we show here that the electronic structure of the Fe-center differs under conditions of $N_2$ reduction reaction ($N_2$RR), $CO_2$ reduction reaction ($CO_2$RR), and HER at deep cathodic potentials, compared to that of bare Fe-center. Further, considering the operating potential ranges of methanol and ethanol oxidation reactions (MOR and EOR) and ORR, it is likely that the nature of the active site will change as the electrode potential increases. Finally, oxygen evolution reaction (OER) will proceed at the oxidized Fe-centers and not the pristine ones. Therefore, we suggest that these effects should be accounted for when modelling electrocatalytic processes on SACs and when interpreting the results of *ex-situ* characterization of SACs.

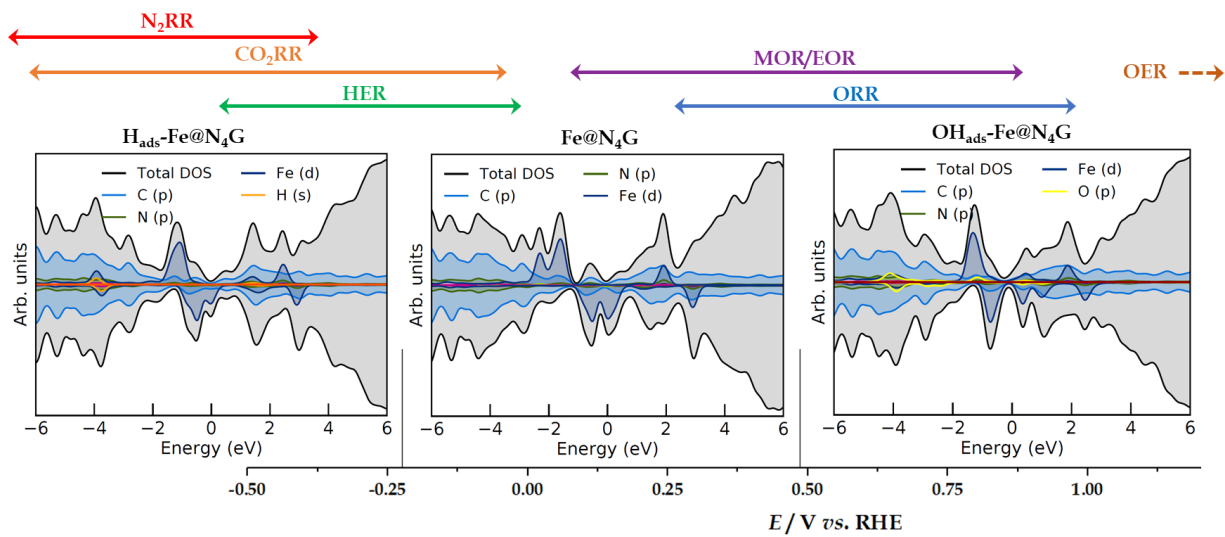

**Figure 6.** Electronic structure of the Fe-center in $Fe@N_4$-graphene SAC in different potential ranges aligned with operative ranges of several electrocatalytic reactions.

One of the main paradigms in SACs is that the tailoring of coordination spheres can be used to tune catalytic activity and selectivity. As one of the examples, one can mention the work on tuning the environment of $Co-N_4$ SACs for enhancing $H_2O_2$ production *via* oxidized nearest carbon neighbors of the $Co-N_4$ moiety [59]. In this work, the oxygen content was controlled during the synthesis and not electrochemically. This is in line with our results (**Figure 7**), suggesting that metal center oxidation is preferred over carbon lattice for metals with lower *d*-band filling. However, the reconstruction of active centers can take place under operating conditions. For example, it was found recently that oxygen reconstitutes the Cu-N$_2$C$_2$ site in Cu-based SACs and that the Cu-N$_2$C$_2$-O site is the actual active species of alkaline ORR [27].

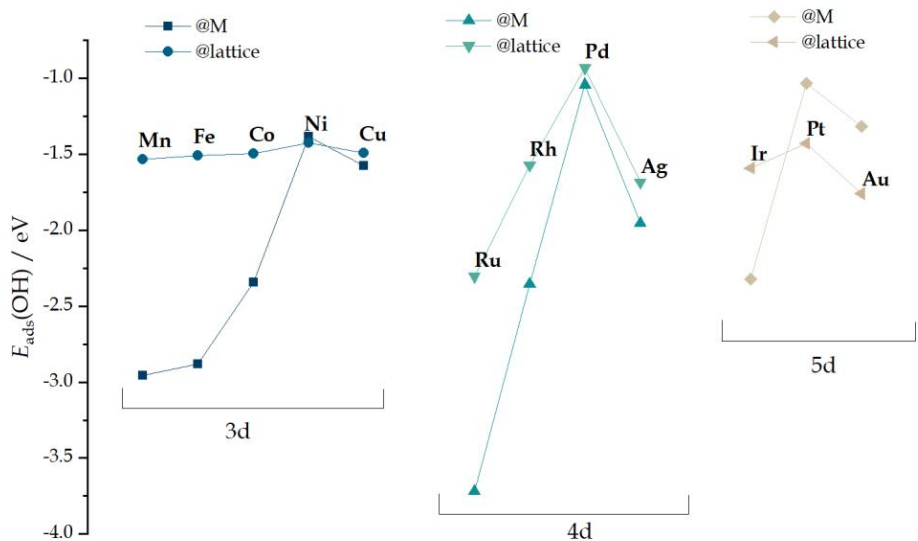

**Figure 7.** Overall trends in OH adsorption on metal sites and carbon lattice adjacent to M-N$_4$ moiety. Looking at the periods of the PTE, metal centers with a lower *d*-band filling strongly prefer OH$_{ads}$ formation, and the differences diminish as the *d*-band filling increases.

While electrochemical processes can cause blockage, alteration of the electronic structure, and restructuring of the local environment of the active sites, it is also necessary to consider the effects of reaction intermediates on the stability of studied SACs. This particularly relates to the effects of intermediately formed H$_2$O$_2$ (in acidic media) or HOO$^-$ if these SACs are used to catalyze ORR. Previously, the effects of H$_2$O$_2$ on the activity and selectivity of Fe-N and Co-N SACs have been confirmed by *ex-situ* catalyst treatment with peroxide [64–66]. The oxidation of the carbon matrix caused the reduction of activity, selectivity, increased oxophilicty of the surface, and led to flooding issues in fuel cells, but the effects were more pronounced in acidic media and partially reversible [64]. In the case of Fe-based catalysts, H$_2$O$_2$ can also undergo Fenton reaction on the Fe centers [67], forming radical species which could easily react with carbon lattice.

Using Fukui indices, here we show for the case of Fe@N$_4$ catalysts (**Figure 8**) that the local reactivity of the carbon lattice is affected by the state of the Fe-center. The carbon sites susceptible to nucleophile or electrophile attack are in both cases located in the third coordination sphere of the Fe@N$_4$ moiety, but the sites suitable for radical attack are located in the first coordination sphere. Moreover, Fe@N$_4$-graphene catalyst with Fe$^{2+}$ center is less susceptible to nucleophile and electrophile attack than the corresponding OH$_{ads}$-Fe@N$_4$-graphene system, which would correspond to the state at high electrode potentials (lower ORR overvoltages). On the other hand, the former system is more susceptible to the radical attack than its oxidized form. When this is combined with the requirement for a relatively low pH and the presence of $Fe^{2+}$ centers for Fenton reaction to generate OH• radicals, it is possible to understand the mentioned effects of $H_2O_2$ treatment in acidic media on the activity of Fe-N catalysts. Namely, we suggest that the effect is due to the reactivity of the carbon lattice plane rendered by the state of the Fe-center combined with the proper state of the Fe-center, enabling radical formation through the Fenton process.

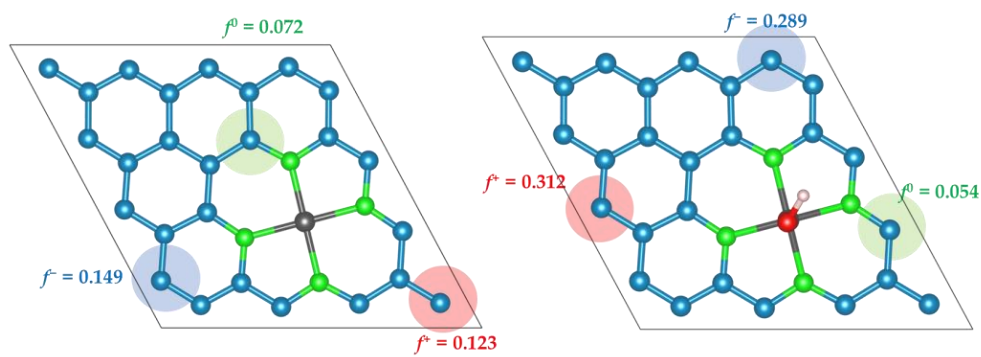

**Figure 8.** Fukui indices for the sites of nucleophile, electrophile, and radical attack on the $M_4$-graphene lattice of Fe@$N_4$-graphene (left) and OH$_{ads}$-Fe@$N_4$-graphene (right).

As discussed so far, the oxidation of the metal centers in M@$N_4$-graphene SACs is omitted when these SACs are modelled to analyze different electrochemical processes. On the other side, in the case of Fe-N catalysts, such oxidation was confirmed experimentally, using *in operando* XANES and FT-EXAFS [23]. However, synchrotron-based techniques are not widely available, and their routine use is still not the case. Thus, the question is whether such a change in the nature of the metal centers can be assessed using more available techniques. Recently, the combination of *in situ* electrochemical IR spectroscopy and DFT calculations has been used to successfully address the local coordination and reactivity of Pt SACs [68]. The results of the mentioned study suggest that *in situ* electrochemical IR spectroscopy has sufficient sensitivity and resolution to be used as a probe for the state of active centers in SACs. Therefore, we have analyzed the vibration spectra of studied SACs, particularly focusing on the OH$_{ads}$-M@$N_4$-graphene systems.

In contrast to the case of pristine M@$N_4$-graphene systems, where the highest frequency vibration corresponds to the in-plane vibrations around the M@$N_4$ moiety (**Figure 9**, left), for OH$_{ads}$-M@$N_4$-graphene systems, the highest frequency vibration is the stretching O–H vibration, which is weakly coupled to the vibrations of the lattice (**Figure 9**, right). The frequency of this vibration is dependent on the specific SAC and the location of the OH group (binding to the metal center of carbon lattice, **Table 3**). We find the O–H vibration positions at relatively high wavenumbers, knowing that O–H vibration for hydroxyl group on carbon lattice is typically located below 3500 cm$^{-1}$ [59,69]. Thus, in principle, it could be possible to use this vibration as the fingerprint of OH$_{ads}$ formation at the metal centers of M@$N_4$ moieties, although it might be difficult to resolve this vibration from a rather intensive hydroxyl group vibration on carbon lattice. However, examples of O–H vibrations on Fe-N SACs at high wavenumbers can be found in literature, although ascribed to the hydroxyls on carbon lattice. In ref. [70], authors reported O–H vibrations on Fe-N SACs at 3689 cm$^{-1}$ using *ex-situ* FTIR, in excellent agreement with our estimate (**Table 3**). In other cases, a fine structure of the FTIR spectra can be seen in the region corresponding to the O–H vibrations [69], suggesting that different OH groups are present in the samples. Another possibility for identifying the presence of the OH phase on metal centers is to use the M–OH vibration, which should be located far from the C–OH vibration, which is in carbons located around 1400 cm$^{-1}$[71]. The shift in the position of this vibration is due to a much larger mass of metal centers than that of the carbon atom. For example, for oxomanganese porphyrin complexes, it should be below 800 cm$^{-1}$ [72]. This possibility is yet to be investigated, but a definite confirmation of the changes of the SACs active site architecture under operating conditions would be of great importance for further improvement of this class of materials in terms of activity and stability.

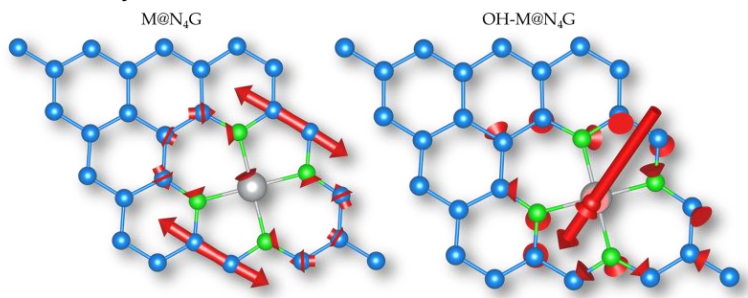

**Figure 9.** The highest wavenumber vibrations modes of M@N$_4$-graphene (left) and OH$_{ads}$-M@N$_4$-graphene (right) SACs presented as vector fields at atomic sites. Visualization is done using the VaspVib2XSF Python code [73].

**Table 3**. OH stretching vibrations (cm$^{-1}$) for studied SACs

| M | OH@M | OH@N$_4$-graphene lattice |
|---|---|---|
| Mn | 3727 | |
| Fe | 3688 | |
| Co | 3706 | |
| Ni | | 3658 |
| Cu | 3732 | |
| Ru | 1412* | |
| Rh | 3691 | |
| Pd | 3713 | |
| Ag | | 3677 |
| Ir | 3684 | |
| Pt | | 3715 |
| Au | | 3708 |

* dissociated OH group

## 4. Conclusions

The DFT results presented in this study confirm that the nature of metal sites in M@N$_4$-graphene SACs (M = Mn, Fe, Co, Ni, Cu, Ru, Rh, Pd, Ag, Ir, Pt, or Au) changes depending on the operating electrode potential and pH. Regarding the influence of electrode potential, the studied SACs exhibit enhanced thermodynamic stability of metal centers towards dissolution compared to bulk metal phases. The stability increases along the periods of PTE, and drops for coinage metals (Cu, Ag, Au). Regarding the influence of pH, while some metals centers (Mn, Co, Ni) are found to dissolve in acidic media, others (Ru, Rh, Ir, Pd, Pt) are expected to be stable at any pH. The location of the center of the main catalytic activity of SACs depends on the electronic configuration of metal used *via* preferential adsorption of electrolyte species. The metals with electron configuration $d^{n\leq 8}$ are found to be the centers of reactivity in M@N$_4$-graphene and the preferential sites for all the investigated adsorbates. The affinity of M@N$_4$-graphene SACs towards H$_{ads}$ and their oxophilicity decrease along the period as H$_{ads}$ formation starts at more negative potentials, while the potentials for OH$_{ads}$ and O$_{ads}$ deposition shift to higher potentials. Depending on the electrode potential, the investigated metals can be bare or covered by H$_{ads}$, O$_{ads}$ or OH$_{ads}$. The mentioned spectator species can block metal sites and induce alteration of the SAC's electronic structure. The possibility of those changes must be considered when dealing with SACs - both in theoretical modelling and interpreting the results of their *ex-situ* characterization.

**Acknowledgment**

Funding: This research was funded by the Science Fund of the Republic of Serbia (PROMIS project RatioCAT); the NATO Science for Peace and Security Programme, grant G5729; the Ministry of Education, Science and Technological Development of the Republic of Serbia (Contract No. 451-03-68/2020-14/200146); and the Carl Tryggers Foundation for Scientific Research (grant no. 18:177), Sweden; and Swedish Research Council (grant no 2019-05580). The computations and data handling were enabled by resources provided by the Swedish National Infrastructure for Computing (SNIC) at the National Supercomputer Centre (NSC) at Linköping University, partially funded by the Swedish Research Council through grant agreement No. 2018-05973.